\documentclass[12pt]{iopart}
\usepackage{latexsym}
\usepackage{amscd}
\begin{document}
\jl{1}
 \def\lambdabar{\protect\@lambdabar}
\def\@lambdabar{%
\relax
\bgroup
\def\@tempa{\hbox{\raise.73\ht0
\hbox to0pt{\kern.25\wd0\vrule width.5\wd0
height.1pt depth.1pt\hss}\box0}}%
\mathchoice{\setbox0\hbox{$\displaystyle\lambda$}\@tempa}%
{\setbox0\hbox{$\textstyle\lambda$}\@tempa}%
{\setbox0\hbox{$\scriptstyle\lambda$}\@tempa}%
{\setbox0\hbox{$\scriptscriptstyle\lambda$}\@tempa}%
\egroup
}

\def\bbox#1{%
\relax\ifmmode
\mathchoice
{{\hbox{\boldmath$\displaystyle#1$}}}%
{{\hbox{\boldmath$\textstyle#1$}}}%
{{\hbox{\boldmath$\scriptstyle#1$}}}%
{{\hbox{\boldmath$\scriptscriptstyle#1$}}}%
\else
\mbox{#1}%
\fi
}
\newcommand{\muv}{\bbox{\mu}}
\newcommand{\mc}{{\mathcal M}}
\newcommand{\pc}{{\mathcal P}}
 \newcommand{\mv}{\bbox{m}}
\newcommand{\pv}{\bbox{p}}
\newcommand{\tv}{\bbox{t}}
\def\msf{\hbox{{\sf M}}}
\def\msft{\bbox{{\sf M}}}
\def\psf{\hbox{{\sf P}}}
\def\psft{\bbox{{\sf P}}}
\def\Nsf{\hbox{{\sf N}}}
\def\Nsft{\bbox{{\sf N}}}
\def\Tsf{\hbox{{\sf T}}}
\def\Tsft{\bbox{{\sf T}}}
\def\Asf{\hbox{{\sf A}}}
\def\Asft{\bbox{{\sf A}}}
\def\Bsf{\hbox{{\sf B}}}
\def\Bsft{\bbox{{\sf B}}}
\def\Lsf{\hbox{{\sf L}}}
\def\Lsft{\bbox{{\sf L}}}
\def\Ssf{\hbox{{\sf S}}}
\def\Ssft{\bbox{{\sf S}}}
\def\Mtens{\bi{M}}
\def\msfsim{\bbox{{\sf M}}_{\scriptstyle\rm(sym)}}
\newcommand{\mcsim}{ {\sf M}_{ {\scriptstyle \rm {(sym)} } i_1\dots i_n}}
\newcommand{\mcs}{ {\sf M}_{ {\scriptstyle \rm {(sym)} } i_1i_2i_3}}

\newcommand{\beqan}{\begin{eqnarray*}}
\newcommand{\eeqan}{\end{eqnarray*}}
\newcommand{\beqa}{\begin{eqnarray}}
\newcommand{\eeqa}{\end{eqnarray}}

 \newcommand{\suml}{\sum\limits}
 \newcommand{\sumd}{\suml_{\mathcal D}}
\newcommand{\intl}{\int\limits}
\newcommand{\rvec}{\bbox{r}}
\newcommand{\xivec}{\bbox{\xi}}
\newcommand{\Avec}{\bbox{A}}
\newcommand{\Rvec}{\bbox{R}}
\newcommand{\Evec}{\bbox{E}}
\newcommand{\Bvec}{\bbox{B}}
\newcommand{\Svec}{\bbox{S}}
\newcommand{\avec}{\bbox{a}}
\newcommand{\nablav}{\bbox{\nabla}}
\newcommand{\nuvec}{\bbox{\nu}}
\newcommand{\bvec}{\bbox{\beta}}
\newcommand{\vvec}{\bbox{v}}
\newcommand{\jvec}{\bbox{j}}
\newcommand{\nvec}{\bbox{n}}
\newcommand{\pvec}{\bbox{p}}
\newcommand{\mvec}{\bbox{m}}
\newcommand{\evec}{\bbox{e}}
\newcommand{\eps}{\varepsilon}
\newcommand{\la}{\lambda}
\newcommand{\rad}{\mbox{\footnotesize rad}}
\newcommand{\scr}{\scriptstyle}
\newcommand{\latens}{\bbox{\Lambda}}
\newcommand{\pitens}{\bbox{\Pi}}
\newcommand{\cm}{{\cal M}}
\newcommand{\cp}{{\cal P}}
\newcommand{\beq}{\begin{equation}}
\newcommand{\eeq}{\end{equation}}
\newcommand{\ptens}{\bbox{\sf{P}}}
\newcommand{\Ptens}{\bbox{P}}
\newcommand{\Ttens}{\bbox{\sf{T}}}
\newcommand{\Ntens}{\bbox{\sf{N}}}
\newcommand{\Ncal}{\bbox{{\cal N}}}
\newcommand{\Atens}{\bbox{\sf{A}}}
\newcommand{\Btens}{\bbox{\sf{B}}}
\newcommand{\dom}{\mathcal{D}}
\newcommand{\al}{\alpha}
\newcommand{\sym}{\scriptstyle \rm{(sym)}}
\newcommand{\Tcal}{\bbox{{\mathcal T}}}
\newcommand{\Nmc}{{\mathcal N}}
\renewcommand{\d}{\partial}
\def\rmi{{\rm i}}
\def\rme{\hbox{\rm e}}
\def\rmd{\hbox{\rm d}}
\newcommand{\ct}{\mbox{\Huge{.}}}
\newcommand{\Laop}{\bbox{\Lambda}}
\newcommand{\Ssfs}{{\scriptstyle \Ssft^{(n)}}}
\newcommand{\Lsfs}{{\scriptstyle \Lsft^{(n)}}}
\newcommand{\psfr}{\widetilde{\psf}}
\newcommand{\msfr}{\widetilde{\msf}}
\newcommand{\msftr}{\widetilde{\msft}}
\newcommand{\psftr}{\widetilde{\psft}}
\newcommand{\qdot}{\stackrel{\cdot\cdot\cdot\cdot}}

\title{ Multipolar radiation and the gauge invariant reduction  
of  multipole tensors}
\author{C. Vrejoiu and Cr. Stoica}
\address{Facultatea de Fizic\v{a}, Universitatea din Bucure\c{s}ti, 077 125, 
Bucure\c{s}ti-M\v{a}gurele, Rom\^{a}nia,  E-mail :  cvrejoiu@yahoo.com   }
\begin{abstract}
Cmpact general formulae for the energy, momentum and angular momentum 
radiated by confined systems of charges and currents are presented in terms of their multipole Cartesian 
tensors. Besides the usual electric and magnetic multipoles, a family of 
toroid (anapole) electromagnetic tensors, as well as some mean squared radii contributions 
are standed out through a method given in previous works of the authors. 
\end{abstract}
\section{Introduction}
    The purpose of the present paper is to obtain general formulae for the energy, momentum 
	and angular momentum loss in the case of a bounded charge and current distribution, in the form 
	of multipolar expansions, using only fully symmetric and trace-free (STF) electric , 
	magnetic and 
	anapole (toroidal) multipole tensors, together with some mean squared radii. The gauge 
	invariant technique of reducing the usual Cartesian electric and magnetic multipole 
	moments to STF ones , presented in \cite{vr} , is used for this end.
    In the section 2 the multipolar expansions of the scalar and vector potentials, $\Phi$ 
	 and $\Avec$, 
	 are presented in terms of electric and magnetic Cartesian multipole moments.
    In section 3 the coresponding expansions of the electric and magnetic fields, $\Evec$
	 and $\Bvec$,  
	are given. The relevant terms for the purposes of this paper are only those that fall 
	off as fast as $1/r$ and $1/r^2$ as $r\to\infty$  given in equations \eref{db1} and 
	\eref{de1}.
    In section 4 the total  radiation intensity (energy loss ) is 
	calculated and given by equation \eref{iepsmu} in the form of an expansion in terms 
	of STF  multipole moments .
	    The same calculation is performed in section 5 and 6 respectively for the total 
	momentum loss and angular momentum loss. In section 7, the approximation's 
	criteria are discussed and, finally, one presents explicit expressions for the radiated 
	energy, momentum and angular momentum including all the electric and magnetic 
	multipoles up to order three (octopoles). These results are compared with some 
	incomplete expressions existing in literature. 
	
    In appendix A one can find the definition formulae for the reduced electric and 
	magnetic multipole moments .
\section{General Formulae}
We shall accept in the  following  a compromise regarding the unit systems used for writing Maxwell's 
equations and the different definitions and equations for electromagnetic theory. Let the 
Maxwell's equations be written as
\beqa\label{Maxwell}
&~& \nablav\times \Bvec=\frac{\mu_0}{\al}\jvec+\frac{\eps_0\mu_0}{\al}\frac{\d\Evec}{\d t},
\;\;\; \nablav\times\Evec=-\frac{1}{\al}\frac{\d\Bvec}{\d t}\nonumber\\
&~& \nablav\Bvec=0,\;\;\; \nablav\cdot\Evec=\frac{1}{\eps_0}\rho
\eeqa   
where the positive parameters $\eps_0,\;\mu_0,\;\al$ are constants depending on the 
system of units. Considering the consequences of Maxwell equations, from physical reasons,  
only two independent constants will be needed because of the constraint
\beqa\label{constr}
&~&\frac{\al^2}{\eps_0\mu_0}=c^2
\eeqa
$c$ standing for the   speed of light in  vacuum.
\par If the Maxwell equations are written in the form above, than  the fields
 $\Evec,\,\Bvec$ and the potentials $\Phi,\,\Avec$ are related by the equations
\beqa\label{FP}
&~&\Bvec=\nablav\times\Avec,\;\;\Evec=-\nablav\Phi-\frac{1}{\al}\frac{\d \Avec}{\d t}
\eeqa
while the retarded potentials are given by
$$\Avec(\rvec,t)=\frac{\mu_0}{4\pi\al}\intl_{\mathcal D}\frac{\jvec(\xivec,t-R/c)}
{R}\rmd\xi^3,\;\;\Phi(\rvec,t)=\frac{1}{4\pi\eps_0}\intl_{\mathcal D}
\frac{\rho(\xivec,t-R/c)}
{R}\rmd\xi^3,\;\; R=\vert\rvec-\xivec\vert.$$

\par   
One can  write the equations and formulae above in the International unit system 
 by taking  
$\al=1$ and the  IS values for $\eps_0,\;\mu_0$.  The same relations can be written  
 in  Heaviside-Lorentz unit system by taking  $ \eps_0=1,\,\mu_0=1,\,\al=c$ 
in cgs units while  in Gauss unit system  $\eps_0=1/4\pi,\,\mu_0=4\pi,\,\al=c$ .   
\par The goal of this compromise is to  compare easily the results from the present paper 
with similar results in different issues written either in IS  or in Gauss or 
Heaviside-Lorentz units.

 Let us consider a charge $\rho(\rvec,t)$ and a current $ \jvec(\rvec,t)$  
distributions having supports included in a finite  domain  $\dom$. Choosing the 
origin $O$ of the Cartesian axes in $\dom$, the retarded  vector and scalar 
potentials at a point outside ${\cal D}$,  $\rvec=x_i\evec_i$,  will be  given by the  
multipolar expansions

\beqa\label{dezv1}
\fl&~&\frac{4\pi}{\mu_0}\Avec(\rvec,t)=\nablav\times\suml_{n\geq 1}\frac{(-1)^{n-1}}
{n!}\nablav^{n-1}\ct\big[\frac{1}{r}\msft^{(n)}(t_0)\big]
+\frac{1}{\al}\suml_{n\geq 1}\frac{(-1)^{n-1}}
{n!}\nablav^{n-1}\ct\big[\frac{1}{r}\dot{\psft}^{(n)}(t_0)\big]\nonumber\\
\fl&~&=\evec_i\eps_{ijk}\d_j\suml_{n\geq 1}\frac{(-1)^{n-1}}{n!}\d_{i_1}\dots\d_{i_{n-1}}
\big[\frac{\msf_{i_1\dots i_{n-1},\,k}(t_0)}{r}\big]\nonumber\\
\fl&~&+\frac{1}{\al}\evec_i\suml_{n\geq 1}\frac{(-1)^{n-1}}{n!}\d_{i_1}\dots\d_{i_{n-1}}
\big[\frac{\dot{\psf}_{i_1\dots i_{n-1}\,i}(t_0)}{r}\big],\;\;\;\;
\;\;\;\;t_0=t-\frac{r}{c},
\\
\fl&~& 4\pi\eps_0\Phi(\rvec,t)=\suml_{n\geq 0}\frac{(-1)^n}{n!}\nablav^n\ct
\big[\frac{\psft^{(n)}(t_0)}{r}\big]
=\suml_{n\geq 0}\frac{(-1)^n}{n!}\d_{i_1}
\dots\d_{i_n}\big[\frac{\psf_{i_1\dots i_n}(t_0)}{r}\big].\nonumber
\eeqa
The electric and magnetic moments are defined as
\beqa\label{defpm}
\fl&~&\psft^{(n)}(t)=\intl_{\dom}\xivec^n\rho(\xivec,t)\rmd^3\xi:\;
\psf_{i_1\dots i_n}=\intl_{\dom}\xi_{i_1}\dots\xi_{i_n}\rho(\xivec,t)\rmd^3\xi\\
\fl&~&\msft^{(n)}(t)=\frac{n}{(n+1)\al}\intl_{\dom}\xivec^n\times\jvec(\xivec,t)\rmd^3\xi:\;\;
\msf_{i_1\dots i_n}=\frac{n}{(n+1)\al}\intl_{\dom}\xi_{i_1}\dots \xi_{i_{n-1}}(\xivec\times\jvec)_{i_n}
\rmd^3\xi\nonumber
\eeqa
 It was shown in \cite{vr}  that we can perform  such transformations of the multipole 
 tensors
\beqa\label{red}
&~&\psft^{(n)}\longrightarrow \widetilde{\psft}^{(n)},\;\;
\msft^{(n)}\longrightarrow \widetilde{\msft}^{(n)},
\eeqa
(where $\widetilde{\psft}^{(n)},\;\widetilde{\msft}^{(n)}$ are STF tensors) so that,  if $\widetilde{\Avec}$ and $\widetilde{\Phi}$ 
 are obtained from equations \eref{dezv1} after the substitutions \eref{red}, the 
 correspondence $(\Avec,\,\Phi)\longrightarrow (\widetilde{\Avec},\,\widetilde{\Phi})$ 
is a gauge transformation. 
\par In Appendix A are presented the principal results of the reducing procedure. In the following 
we will suppose that  all the multipole moments are represented by  reduced STF  
tensors.

\section{Expansions of the fields $\Evec$ and $\Bvec$}
Let us write the multipole expansion of $\Bvec=\nablav\times\Avec$. Use of equation 
\eref{dezv1} and formulae 
$\nablav\times(\nablav\times \bbox{a})=\nablav\cdot(\nablav\cdot\bbox{a})-\Delta\,\bbox{a}$ 
and $$\big[\Delta-\frac{1}{c^2}\frac{\d^2}{\d t^2}\big]\frac{f(t-r/c)}{r}=0,\;
\mbox{for}\;\; r\neq 0,$$
leads to the expansion
\beqa\label{dezvB}
&~&\fl\Bvec(\rvec,t)=\frac{\mu_0}{4\pi}\suml_{n\geq 1}\frac{(-1)^{n-1}}{n!}
\nablav\cdot\big[\nablav^{n}\ct\big[\frac{1}{r}\msftr^{(n)}(t_0)\big]\big]\nonumber\\
&~&\fl-\frac{\mu_0}{4\pi c^2}\suml_{n\geq 1}\frac{(-1)^{n-1}}{n!}\nablav^{n-1}\ct\big[\frac{1}{r}
\ddot{\msftr}^{(n)}(t_0)\big]
+\frac{\mu_0}{4\pi\al}\nablav\times\suml_{n\geq 1}\frac{(-1)^{n-1}}{n!}
\nablav^{n-1}\ct\big[\frac{1}{r}\dot{\psfr}^{(n)}(t_0)\big],\nonumber\\
&~&\fl t_0=t-\frac{r}{c}.
\eeqa
For $\Evec$ we get
\beqa\label{dezvE}
\fl&~&\Evec(\rvec,t)=-\frac{\al}{4\pi\eps_0 c^2}\nablav\times\suml_{n\geq 1}
\frac{(-1)^{n-1}}{n!}\nablav^{n-1}\ct\big[\frac{1}{r}\dot{\msftr}^{(n)}(t_0)\big]\nonumber\\
\fl&~&-\frac{1}{4\pi\eps_0c^2}\suml_{n\geq 1}\frac{(-1)^{n-1}}{n!}\nablav^{n-1}\ct
\big[\frac{1}{r}\ddot{\psfr}^{(n)}(t_0)\big]-\frac{1}{4\pi\eps_0}\nablav\suml_{n\geq 1}
\frac{(-1)^{n-1}}{n!}\nablav^{n}\ct\big[\frac{1}{r}\psfr^{(n)}(t_0)\big]
\eeqa

\par In order to obtain the behavior of the fields $\Evec$ and $\Bvec$ at large distances, 
 we need the formula  for the partial derivatives of a retarded arbitrary solution 
 of the wave equation. The following formula can be easily obtained:
 \beqa\label{dezvf}
  &~&\fl\d_{i_1}\dots\d_{i_n}\big[\frac{1}{r}f(t_0)\big]=\frac{1}{r}
 \frac{(-1)^n}{c^n}\nu_{i_1}\dots\nu_{i_n}\frac{\rmd^nf(t_0)}{\rmd t^n}\nonumber\\
&~& \fl +\frac{1}{r^2}\frac{(-1)^n}{c^{n-1}}\left[D_n\,\nu_{i_1}\dots\nu_{i_n}
 \frac{\rmd^{n-1}f(t_0)}{\rmd t^{n-1}} - \nu_{\{i_1}\dots
 \nu_{i_{n-2}}\delta_{i_{n-1}i_n\}} \frac{\rmd^{n-1}f(t_0)}{\rmd t^{n-1}}\right]+O(\frac{1}{r^3})
 \eeqa
where
\beqa\label{relrec}
&~&\fl D_n=D_{n-1}+n,\;D_0=0,\;\;\;\nuvec=\frac{\rvec}{r}.
\eeqa
In  equation \eref{dezvf} and in the following  we understand by $A_{\{i_1\dots i_n\}}$ 
the sum over all the permutations of the symbols $i_q$  that give distinct terms.  
The use of the formula \eref{dezvf} in equations \eref{dezvB} and \eref{dezvE}, and also 
of the well known properties of STF tensors, allows the ordering of the contributions 
in the expansions above as a $1/r$ power series.

\beqa\label{db1}
&~&\fl \Bvec(\rvec,t)=\frac{\mu_0}{4\pi r}\suml_{n\geq 1}\frac{1}{n!c^{n+1}}\left\{
\big[\nuvec^n\ct\msftr^{(n)}_{,\,n+1}\big]\nuvec -
\big[\nuvec^{n-1}\ct\msftr^{(n)}_{,\,n+1}\big]-\frac{c}{\al}\nuvec\times\big[
\nuvec^{n-1}\ct\psfr^{(n)}_{,\,n+1}\big]\right\}\nonumber\\
&~&\fl+\frac{\mu_0}{4\pi r^2}\suml_{n\geq 1}\frac{1}{n!c^n}\left\{
D_{n+1}\big[\nuvec^n\ct\msftr^{(n)}_{,\,n}\big]\nuvec-
D_n\big[\nuvec^{n-1}\ct\msftr^{(n)}_{,\,n}\big]
-\frac{cD_n}{\al}\,\nuvec\times\big[\nuvec^{n-1}
\ct\psf^{(n)}_{,\,n}\big]
\right\}+\dots
\eeqa
with the notation $f_{,\,n}=\rmd^nf(t_0)/\rmd t^n$ and
\beqa\label{de1}
&~&\fl\Evec(\rvec,t)=\frac{1}{4\pi\eps_0r}\suml_{n\geq 1}\frac{1}{n!c^{n+1}}\left\{
\big[\nuvec^n\ct\psfr^{(n)}_{,\,n+1}\big]\nuvec
-\big[\nuvec^{n-1}\ct\psfr^{(n)}_{,\,n+1}\big]+
\frac{\al}{c}\nuvec\times\big[\nuvec^{n-1}\ct\msftr^{(n)}_{,\,n+1}\big]
\right\}\nonumber\\
&~&\fl +
\frac{1}{4\pi\eps_0r^2}\suml_{n\geq 1}\frac{1}{c^nn!}\left\{
D_{n+1}\big[\nuvec^n\ct\psfr^{(n)}_{,\,n}\big]
\nuvec-D_{n}\big[\nuvec^{n-1}\ct\psfr^{(n)}_{,\,n}\big]+
\frac{\al }{c}D_n
\nuvec\times\big[\nuvec^{n-1}\ct\msftr^{(n)}_{,\,n}\big]\right\}+\dots
\eeqa
Let us write $\Evec=\Evec^{(1)}+\Evec^{(2)}+O(1/r^3)$   where $\Evec^{(1)}$ and $\Evec^{(2)}$ are 
respectively, proportional with $1/r$ and $1/r^2$ and, similarly, $\Bvec=\Bvec^{(1)}+
\Bvec^{(2)}+O(1/r^3)$. The parts $\Evec^{(1)}$ and $\Bvec^{(1)}$ of the field are 
purely  transverse fields, satisfying the properties
\beqa\label{transv}
&~&\fl\nuvec\cdot\Evec^{(1)}=0,\;\nuvec\cdot\Bvec^{(1)}=0;\;\;\Evec^{(1)}=\frac{c}{\al}
\Bvec^{(1)}\times\nuvec,\;,\;\;
\eps_0\vert\Evec^{(1)})\vert^2=\frac{1}{\mu_0}\vert\Bvec^{(1)}\vert^2.
\eeqa 
\section{Radiation Intensity}
The Poynting vector will be written  in terms of $\Evec^{(1)}$ and $\Bvec^{(1)}$ by  
considering the equations \eref{transv}:
\beqan
\bbox{S}=\frac{\al}{\mu_0}\left(\Evec\times\Bvec\right)
=\eps_0\vert\Evec^{(1)}\vert^2\,c\nuvec+O(\frac{1}{r^3})=
\frac{1}{\mu_0}\vert\Bvec^{(1)}\vert^2c\nuvec+O(\frac{1}{r^3})
\eeqan
The total  radiated power may be written as the limit of the integral on the sphere of 
radius $r$ as $r\to \infty$:
\beqa\label{tp}
\fl&~&{\mathcal I}=\lim_{r\to\infty}\int r^2\nuvec\cdot\bbox{S}\,\rmd\Omega(\nuvec)=
\int r^2\nuvec\cdot\bbox{S}\,\rmd\Omega(\nuvec)=\frac{cr^2}{\mu_0}\int\vert\Bvec^{(1)}\vert
^2\,\rmd\Omega(\nuvec)\nonumber\\
\fl&~&=\frac{4\pi c}{\mu_0}<r^2\vert\Bvec^{(1)}\vert^2>
\eeqa
 where $$<f(\nu)>=\frac{1}{4\pi}\int f(\nu)\,\rmd\Omega(\nuvec).$$
We obtain from equation \eref{db1} that
\beqa\label{bsq}
\fl&~&r^2\vert\Bvec^{(1)}\vert^2=\big(\frac{\mu_0}{4\pi}\big)^2\suml_{n,m\geq 1}
\frac{1}{n!m!}\left\{-\frac{1}{c^{n+m+2}}\big(\nuvec^n\ct\msftr^{(n)}_{,\,n+1}\big)
\big(\nuvec^m\ct\msftr^{(m)}_{,\,m+1}\big)\right.\nonumber\\
\fl&~&\left.+\frac{1}{c^{n+m+2}}
\big(\nuvec^{n-1}\ct\msftr^{(n)}_{,\,n+1}\big)\ct\big(\nuvec^{m-1}
\ct\msftr^{(m)}_{,\,m+1}\big)+\frac{1}{\al c^{n+m+1}}\big(\nuvec^{n-1}\ct
\msftr^{(n)}_{,\,n+1}\big)\ct\big[\nuvec\times\big(\nuvec^{m-1}\ct\psftr^{(m)}_{,\,m+1}
\big)\big]\right.\nonumber\\
\fl&~&+\frac{1}{\al c^{n+m+1}}\big[\nuvec\times\big(\nuvec^{n-1}\ct\psftr^{(n)}_{,\,n+1}
\big)\big]\ct  \big(\nuvec^{m-1}\ct\msftr^{(m)}_{,\,m+1}\big)
\nonumber\\
\fl&~&\left.+\frac{1}{\al^2 c^{n+m}}\big(\nuvec^{n-1}\ct\psftr^{(n)}_{,\,n+1}\big)\ct
\big(\nuvec^{m-1}\ct\psftr^{(m)}_{,\,m+1}\big)
-\frac{1}{\al^2 c^{n+m}}\big(\nuvec^{n}\ct\psftr^{(n)}_{,\,n+1}\big)
\big(\nuvec^{m}\ct\psftr^{(m)}_{,\,m+1}\right\}.
\eeqa
 The following  formula is used to perform the average operations, \cite{Thorne}   
\beqa\label{avfor}
  <\nu_1\dots\nu_n>=\left\{\begin{array}{l}0,\;\;n=2k+1,\; \\
  \frac{1}{(2k+1)!!}\delta_{\{i_1i_2}\dots\delta_{i_{n-1}i_n\}},\;n=2k,
  \;\;\;k=0,1,2,\dots\end{array}\right\}
\eeqa
If $\Asft^{(n)},\;\Bsft^{(m)}$ are STF tensors, one can show that 
\beqa\label{plus}
\fl&~&<(\nuvec^k\ct\Asft^{(n)})\ct(\nuvec^{k'}\ct\Bsft^{(m)})>=
\frac{k!\delta_{kk'}}{(2k+1)!!}\left(\Asft^{(n)}\ct \Bsft^{(m)}
\right)
\eeqa
For the  calculation of the averaged square of the vector $\Bvec^{(1)}$ one can 
 apply the following formulae obtained from equation \eref{plus}:
\beqa\label{av1}
\fl&~&<\big(\nuvec^n\ct\msftr^{(n)}\big)\big(\nuvec^m\ct\msftr^{(m)}\big)>=
\frac{n!\delta_{nm}}{(2n+1)!!}\big(\msftr^{(n)}\ct\msftr^{(n)}\big),\; \nonumber\\
\fl&~&<\big(\nuvec^{n-1}\ct\msftr^{(n)}\big)\big(\nuvec^{m-1}\ct\msftr^{(m)}\big)>=
\frac{(n-1)!\delta_{nm}}{(2n-1)!!}\big(\msftr^{(n)}\ct\msftr^{(n)}\big)\nonumber\\
\fl&~&\left<\big(\nuvec^{n-1}\ct\msftr^{(n)}\big)\ct\big[\nuvec\times\big(\nuvec^{m-1}
\ct\psftr^{(m)}\big)\big]\right>=\left<\big(\nuvec^{n-1}\ct\msftr^{(n)}\big)_i
\eps_{ijk}\nu_j\big(\nuvec^{m-1}\ct\psftr^{(m)}\big)_k\right>\nonumber\\
\fl&~&=\eps_{ijk}<\nu_{i_1}\dots i_{n-1}\nu_j\nu_{j_1}\dots \nu_{j_{m-1}}>
\msfr_{i_1\dots i_{n-1}i}\psfr_{j_1\dots j_{m-1}k}=0.
\eeqa

By introducing the expansion \eref{db1} of $\Bvec$  in equation 
\eref{tp}, and using the relations \eref{av1}, one obtains 
\beqa\label{iepsmu}
\fl&~&{\mathcal I}=\frac{\al^2}{4\pi\eps_0c^3}
\suml_{n\geq 1}\frac{n+1}{nn!(2n+1)!!c^{2n}}\left[\big(\msftr^{(n)}_{,\,n+1}\ct
\msftr^{(n)}_{,\,n+1}\big)+ 
\frac{c^2}{\al^2}\big(\psftr^{(n)}_{,\,n+1}\ct\psftr^{(n)}_{,\,n+1}\big)\right]
\eeqa

\section{Recoil Force}
By considering the momentum current density tensor as
\beqa\label{crimp}
\fl&~ {\mathcal T}_{ij}=\frac{1}{2}\big(\eps_0\Evec^2+\frac{1}{\mu_0}\Bvec^2\big)
\delta_{ij}-\big(\eps_0E_iE_j+\frac{1}{\mu_0}B_iB_j\big),
\eeqa
then the recoil force is given by
\beqa\label{rf}
\fl&~&\bbox{F}_R=-\lim_{r\to\infty}\oint_{\Sigma_r}\big(\nuvec\ct{\mathcal T})\rmd S=
-\lim_{r\to\infty}r^2\int\nu_i{\mathcal T}_{ij}\rmd\Omega(\nuvec)\,\evec_j\nonumber\\
\fl&~&=-\lim_{r\to\infty}r^2\int\frac{1}{2}\big(\eps_0\Evec^2+\frac{1}{\mu_0}\Bvec^2\big)
\nuvec\rmd\Omega(\nuvec)=-\frac{4\pi}{\mu_0}\big<r^2\vert\Bvec^{(1)}\vert^2\nuvec\big>.
\eeqa
Let us introduce the expansion \eref{bsq} of $\vert\Bvec^{(1)}\vert^2$ 
in equation \eref{rf} and consider the relations
\beqa\label{av5}
\fl&~&\left<\big(\nuvec^n\ct\msftr^{(n)}_{,\,n+1}\big)\big(\nuvec^m\ct
\msftr^{(m)}_{,\,m+1}\big)\nuvec\right>\nonumber\\
\fl&~&=\frac{n!\delta_{m,n-1}}{(2n+1)!!}\big(\msftr^{(n)}_{,\,n+1}\ct
\msftr^{(n-1)}_{,\,n}\big)
+\frac{(n+1)!\delta_{m,n+1}}{(2n+3)!!}\big(\msftr^{(n)}_{,\,n+1}\ct
\msftr^{(n+1)}_{,\,n+2}\big),\nonumber\\
\fl&~&\left<\big(\nuvec^{n-1}\ct\msftr^{(n)}_{,\,n+1}\big)\big(\nuvec^{m-1}\ct
\msftr^{(m)}_{,\,m+1}\big)\nuvec\right>\nonumber\\
\fl&~&=\frac{(n-1)!\delta_{m,n-1}}{(2n-1)!!}\big(\msftr^{(n)}_{,\,n+1}\ct
\msftr^{(n-1)}_{,\,n}\big)
+\frac{n!\delta_{m,n+1}}{(2n+1)!!}\big(\msftr^{(n)}_{,\,n+1}\ct
\msftr^{(n+1)}_{,\,n+2}\big),\nonumber\\
\fl&~&\big<\big(\nuvec^{n-1}\ct\msftr^{(n)}\big)\ct\big[\nuvec\times\big(\nuvec^{m-1}
\ct\psftr^{(m)}\big)\big]\,\nuvec\big>=
\big<\nu_{i_1}\dots\nu_{i_{n-1}}\nu_i\nu_j\nu_{j_1}\dots\nu_{j_{m-1}}\big>\nonumber\\
\fl&~&\times \evec_i\eps_{qjk}\msfr_{i_1\dots i_{n-1}q}\psfr_{j_1\dots j_{m-1}k}
=-\frac{(n-1)!}{(2n+1)!!}\evec_i\eps_{iqk}\msfr_{i_1\dots i_{n-1}q}
\psfr_{i_1\dots i_{m-1}k}\delta_{nm}\nonumber\\
\fl&~&=-\frac{(n-1)!}{(2n+1)!!}\evec_i\eps_{ijk}\big(\msftr^{(n)}\ct n-1\ct\psftr^{(n)}
\big)_{jk}\delta_{nm},
\eeqa
with the obvious notation,
$$ \big(\msftr^{(n)}\ct n-1\ct\psftr^{(n)}\big)_{jk}=\msf_{i_1,\dots i_{n-1}j}
\psf_{i_1\dots i_{n-1}k}.$$
After observing that the  terms that give nonzero contributions are only those 
containing a $\delta$ factor, one gets  
\beqa\label{Fr}
\fl&~&\bbox{F}_R=-\frac{\mu_0}{4\pi}\suml_{n\geq 1}\frac{1}{c^{2n}(2n+1)!!}\nonumber\\
\fl&~&\times\left\{
\frac{n+1}{cn!}\big(\msftr^{(n)}_{,\,n+1}\ct\msftr^{(n-1)}_{,\,n}\big)+\frac{n+2}{
c^3(2n+3)!!(n+1)!}\big(\msftr^{(n)}_{,\,n+1}\ct\msftr^{(n+1)}_{,\,n+2}\big)\right.\nonumber\\
\fl&~&-\frac{2}{\al cn!n}\evec_i\eps_{ijk}\big(\msftr^{(n)}_{,\,n+1}\ct n-1\ct\psftr^{(n)}
_{,\,n+1}\big)_{jk}+\frac{c(n+1)}{\al^2n!}\big(\psftr^{(n)}_{,\,n+1}\ct
\psftr^{(n-1)}_{,\,n}\big)\nonumber\\
\fl&~&\left.+\frac{n+2}{\al^2c(2n+3)!!(n+1)!}\big(\psftr^{(n)}_{,\,n+1}\ct
\psftr^{(n+1)}_{,\,n+2}\big)\right\}
\eeqa
Since $\msf^{(0)}=0$, and the total electric charge in ${\mathcal D}$ is constant  
($\rmd \psf^{(0)}/\rmd t=0$), then the following formula can be obtained after performing 
some changes of summation indices:
\beqa\label{Frf}
\fl&~&\bbox{F}_R=-\frac{\mu_0}{2\pi c^3}\suml_{n\geq 1}\frac{1}{c^{2n}}
\left\{
\frac{n+2}{(n+1)!(2n+3)!!}\big[\msftr^{(n)}_{,\,n+1}\ct\msftr^{(n+1)}_{,\,n+2}
+\frac{c^2}{\al^2}\big(\psftr^{(n)}_{,\,n+1}\ct\psftr^{(n+1)}_{,\,n+2}\big]\right.\nonumber\\
\fl&~&\left.-\frac{c^2}{\al} \frac{1}{n!n(2n+1)!!}\evec_i\eps_{ijk}
\big(\msftr^{(n)}_{,\,n+1}\ct n-1\ct\psftr^{(n)}_{,\,n+1}\big)_{jk}\right\}
\eeqa

\section{Angular Momentum Loss}
The total angular momentum lost per unit time by  a radiating system is given by the 
flux of the radiated electromagnetic angular momentum through the spherical surface 
of radius $r\to\infty$. The $2$nd-order antisymmetric tensor associated with the 
 electromagnetic angular momentum density 
is defined by its Cartesian components  $m_{ij}=x_ig_j-x_jg_i$ where the electromagnetic  
momentum  density vector is $\bbox{g}=(\eps_0/\al)(\Evec\times\Bvec)$. The angular 
momentum density current is associated with  the $3$rd-order tensor $\bbox{\mu}^{(3)}$; 
$\mu_{kij}=x_i{\mathcal T}_{jk}-x_j{\mathcal T}_{ik}$ with ${\mathcal T}_{jk}$ given 
by equation \eref{crimp}. By introducing the angular momentum density pseudovector 
$\bbox{L}$ with the components $L_i=(1/2)\eps_{ijk}m_{jk}$, and the $2$nd-order 
angular momentum current tensor $\mu^{(2)}_{kl}=(1/2)\eps_{lij}\mu^{(3)}_{kij}$, 
we obtain that       
$$\nu_k\mu^{(2)}_{ki}=-r\left[\eps_0(\nuvec\ct\Evec)(\nuvec\times\Evec)_i
+\frac{1}{\mu_0}(\nuvec\ct\Bvec)(\nuvec\times\Bvec)_i\right]$$
and further that 
\beqa\label{momrad}
\fl\left(\frac{\rmd\bbox{L}}{\rmd t}\right)_{\scriptstyle \mbox{rad}}
&=&-\lim_{r\to\infty}\oint_{\Sigma_r}\nu_k\mu^{(2)}_{ki}\rmd S\nonumber\\
\fl&=&4\pi r^3\left<\frac{1}{\mu_0}\big(\nuvec\ct\Bvec^{(2)}\big)\big(\nuvec\times\Bvec^{(1)}
\big)+\eps_0\big(\nuvec\ct\Evec^{(2)}\big)\big(\nuvec\times\Evec^{(1)}\big)\right>.
\eeqa
By using equations \eref{relrec}-\eref{de1}, one obtains
\beqa\label{nub2}
&~&\fl r^2\,\nuvec\ct\Bvec^{(2)}=frac{\mu_0}{4\pi}\suml_{n\geq 1}\frac{n+1}{n!\,c^n}\big(
\nuvec^n\ct\msftr^{(n)}_{,\,n}\big),
\eeqa
\beqa\label{nue2}
&~&\fl r^2\,\nuvec\ct\Evec^{(2)}=\frac{1}{4\pi\eps_0}\suml_{n\geq 1}\frac{n+1}{n!\,c^n}\big(
\nuvec^n\ct\psftr^{(n)}_{,\,n}\big),
\eeqa
and
\beqa\label{nub1}
&~&\fl r\nuvec\times\Bvec^{(1)}=\frac{\mu_0}{4\pi\al}\suml_{n\geq 1}\frac{1}{c^nn!}\left\{
-\frac{\al}{c}\nuvec\times\big(\nuvec^{n-1}\ct\msftr^{(n)}_{,\,n+1}\big)
-\big(\nuvec^n\ct\psftr^{(n)}_{,\,n+1}\big)\nuvec+\big(\nuvec^{n-1}\ct\psftr^{(n)}_{,\,n+1}
\big)\right\},
\eeqa
\beqa\label{nue1}
&~&\fl r\nuvec\times\Evec^{(1)}=\frac{\mu_0}{4\pi\al}\suml_{n\geq 1}\frac{1}{c^nn!}
\left\{-\frac{c}{\al}\nuvec\times\big(\nuvec^{n-1}\ct\psftr^{(n)}_{,\,n+1}\big)+
\big(\nuvec^n\ct\msftr^{(n)}_{,\,n+1}\big)\nuvec-\big(\nuvec^{n-1}\ct\msftr^{(n)}
_{,\,n+1}\right\}.
\eeqa
We than have the following substitution rules:
\beqa\label{subst}
 \left(\begin{array}{l}\msftr^{(n)}\rightarrow\frac{c^2}{\al^2}\psftr^{(n)}\\
\psftr^{(n)}\rightarrow -\msftr^{(n)}\end{array}\right)\Rightarrow
\left(\begin{array}{l}\nuvec\ct\Bvec^{(2)}\rightarrow \nuvec\ct\Evec^{(2)}\\
\nuvec\times\Bvec^{(1)}\rightarrow \nuvec\times\Evec^{(2)}\end{array}\right).
\eeqa
After using equations \eref{av5} and
\beqa\label{avu}
&~&\fl\big<\big(\nuvec^n\ct\msftr^{(n)}_{,\,n}\big)\big(\nuvec^{m-1}\ct
\psftr^{(m)}_{,\,m+1}\big)\big>=\frac{n!\delta_{m,n+1}}{(2n+1)!!}\big(\msftr^{(n)}_{,\,n}\ct
\psftr^{(n+1)}_{,\,n+2}\big)
\eeqa
and the substitution rules \eref{subst}, we obtain
\beqa\label{medb}
\fl&~&4\pi r^3\big<(\nuvec\ct\Bvec^{(2)})(\nuvec\times\Bvec^{(1)})\big>=
\frac{\mu_0^2}{4\pi}\suml_{n\geq 1}\left[-\frac{n+1}{n!(2n+1)!!\,c^{2n+1}}\evec_i\eps_{ijk}
\big(\msftr^{(n)}_{,\,n}\vert n-1\vert\msftr^{(n)}_{,\,n+1}\big)_{jk}\right.\nonumber\\
\fl&~&\left.-\frac{n+1}{(n-1)!(2n+1)!!\al c^{2n-1}}\big(\msftr^{(n)}_{,\,n}\ct\psftr^{(n-1)}
_{,\,n}\big)+\frac{n+2}{n!(2n+3)!!\al c^{2n+1}}\big(\msftr^{(n)}_{,\,n}\ct\psftr^{(n+1)}
_{,\,n+2}\big)\right],
\eeqa
and
\beqa\label{mede}
\fl&~&4\pi r^3\big<(\nuvec\ct\Evec^{(2)})(\nuvec\times\Evec^{(1)})\big>=
\frac{\mu_0^2}{4\pi}\suml_{n\geq 1}\left[-\frac{n+1}{n!(2n+1)!!\al^4c^{2n-3}}\evec_i\eps_{ijk}
\big(\psftr^{(n)}_{,\,n}\vert n-1\vert\psftr^{(n)}_{,\,n+1}\big)_{jk}\right.\nonumber\\
\fl&~&\left.+\frac{n+1}{(n-1)!(2n+1)!!\al^3 c^{2n-3}}\big(\psftr^{(n)}_{,\,n}\ct\msftr^{(n-1)}
_{,\,n}\big)-\frac{n+2}{n!(2n+3)!!\al^3 c^{2n-1}}\big(\psftr^{(n)}_{,\,n}\ct\msftr^{(n+1)}
_{,\,n+2}\big)\right]
\eeqa
By introducing equations \eref{medb} and \eref{mede} in equation \eref{momrad}, we obtain]
\beqa\label{finmom}
\fl&~&\left(\frac{\rmd\bbox{L}}{\rmd t}\right)_{\scriptstyle \mbox{rad}}=\frac{\mu_0}{4\pi}
\suml_{n\geq 1}\left[-\frac{n+1}{c^{2n+1}n!(2n+1)!!}\evec_i\eps_{ijk}\big(\msftr^{(n)}_{,
\,n}\ct n-1\ct\msftr^{(n)}_{,\,n+1}\big)_{jk}\right.\nonumber\\
\fl&~&-\frac{n+1}{\al c^{2n-1}(n-1)!(2n+1)!!}\big(\msftr^{(n)}_{,\,n}\ct\psftr^{(n-1)}_{,\,n}\big)
+\frac{n+2}{\al c^{2n+1}n!(2n+3)!!}\big(\msftr^{(n)}_{,\,n}\ct\psftr^{(n+1)}_{,\,n+2}\big)
\nonumber\\
\fl&~&-\frac{n+1}{\al^2 c^{2n-1}n!(2n+1)!!}\evec_i\eps_{ijk}\big(\psftr^{(n)}_{,\,n}\ct
n-1\ct\psftr^{(n)}_{,\,n+1}\big)_{jk}\nonumber\\
\fl&~&\left.+\frac{n+1}{\al c^{2n-1}(n-1)!(2n+1)!!}
\big(\psftr^{(n)}_{,\,n}\ct\msftr^{(n-1)}_{,\,n}\big)
-\frac{n+2}{\al c^{2n+1}n!(2n+3)!!}\big(\psftr^{(n)}_{,\,n}\ct\msftr^{(n+1)}_{,\,n+2}\big)
\right]
\eeqa
and, after some algebraic operations, we obtain
\beqa\label{finmomf}
\fl&~&\frac{\rmd\bbox{L}}{\rmd t} =\frac{\mu _{0}}{4\pi c}\sum\limits_{n\geq 1}%
\frac{1}{c^{2n}}\left\{- \frac{n+1}{n!(2n+1)!!}\vec{e}_{i}\varepsilon
_{ijk}\times \right.\nonumber \\
\fl&~&\times \left[ \frac{c^{2}}{\alpha ^{2}}\left( \psftr_{,n}^{\left( n\right)
}\ct n-1\ct \psftr_{,n+1}^{\left( n\right) }\right) _{jk}+\left(
\msftr_{,n}^{\left( n\right) }\ct n-1\ct \msftr_{,n+1}^{\left( n\right)
}\right) _{jk}\right] + \\
\fl&~&+\frac{n+2}{\al n!(2n+3)!!}\left[ \psftr_{,n+1}^{\left( n+1\right)
}\cdot \msftr_{,n+1}^{\left( n\right) }-\msftr_{,n+1}^{\left( n+1\right) }\cdot
\psftr_{,n+1}^{\left( n\right) }
 \left. +\msftr_{,n}^{\left( n\right) }\cdot \psftr_{,n+2}^{\left( n+1\right)
}-\psftr_{,n}^{\left( n\right) }\cdot \msftr_{,n+2}^{\left( n+1\right) }\right]
\right\}.\nonumber
\eeqa

\section{Discussion and conclusions}
The procedure, given in the present paper, for calculating the energy, momentum and 
angular momentum radiated by  arbitrary sources is characterized  by the 
mathematical simplicity. In our calculations only simple algebraic manipulation 
is necessary. The results expressed in different coordinate systems, for example spherical, 
may be obtained by adequate transformations. 
Complete and correct multipole analysis is given in some fundamental 
works as \cite{dub,ra}. In these works, the multipole expansion formulae are introduced 
using the canonical basis of  the solutions of the scalar wave equation of 
Helmholtz. Although the results obtained by this formalism represent an exact description 
of the multipolar expansion applied to the electromagnetic field of arbitrary sources, 
the formulae are rather cumbersome. Due to the complexity of these formulae, the possibility  
 to obtain  erroneous results exists even in the case one calculates the contributions of the 
 low order multipoles to the radiated field. This is the reason why we believe it is necessary 
to give corrections for the radiation formulae obtained in \cite{dub,ra} in addition to 
the corrections given in these papers to the results from \cite{land,jack}.            

\par By applying the formulae for the radiated quantities, some care is necessary 
for obtaining a correct grouping of the different multipolar terms in a given 
approximation. For understanding this problem it is sufficient to refer to the harmonic 
time variation of the charges and currents. Let us consider
$$ \rho(\rvec,t)=\rho_0(\rvec)\rme^{-\rmi\omega t},\;\;\jvec(\rvec,t)=\jvec_0(\rvec)
\rme^{-\rmi\omega t}.$$
By considering the magnetic multipole tensor $\msft(t)$, the $n$-order time derivative 
may be written as
\beqa\label{dnmdtn}
&~&\fl\frac{\rmd^n}{\rmd t^n}\msft^{(n)}=\frac{(-\rmi\omega)^n\,n}{(n+1)\al}
\intl_{\dom}\xivec^n\times\jvec_0(\xivec)\rmd^3\xi\sim \omega^nd^n
\eeqa
where $d$ is the linear dimension of the domain $\dom$. By introducing the wave length 
of the emited radiation, $\lambda=2\pi c/\omega$, we may write
\beqa\label{Mdl}
&~&\fl\frac{1}{c^n}\frac{\rmd^n}{\rmd t^n}\msft^{(n)}\sim \left(\frac{d}{\lambda}\right)^n.
\eeqa
The parameters $\omega$ and $d$ are related, as order of magnitude, to the 
charges velocities, $\omega d\sim v$, such that $\omega d/c\sim v/c$.
In what concerns the electric multipole's contributions, we notice that the contribution 
of the time derivative of the $n$-order electric multipole is of the same order of 
magnitude as the contribution of the $(n-1)th$ order magnetic multipole because 
$$\frac{\rmd}{\rmd t}\psf_{i_1\dots i_n}(t)=\intl_{\dom}\jvec\cdot\nablav (\xi_{i_1}\dots 
\xi_{i_n})\rmd^3\xi=\intl_{\dom}\xi_{\{i_1}\dots\xi_{i_{n-1}}j_{i_n\}}\rmd^3\xi.$$
  Consequently, as is pointed out also in \cite{DV}, in the case of the radiation 
  intensity,  in equations \eref{iepsmu}, \eref{Frf} and \eref{finmomf},  a 
  consistent expansion  should take into account, alongside with the magnetic multipoles up to 
  a given  order  $\mu$,  the electric multipoles up to the order $\eps=\mu+1$.
\par For a comparison with some results from the literature, for example \cite{dub,ra},  
we give below the results of the expansions of the total radiated power, recoil force and 
angular  momentum for $(\mu,\eps)=(4,5)$ pointing out the criteria for grouping the different 
multipolar terms for a given approximation.
\par In order to understand this problem, we give for the case of the radiated power
 a detailed description 
of the procedure used. We denote by ${\mathcal I}^{(\mu,\eps)}$ the radiated power 
obtained by the recursive procedure begining from the order $\mu$ for the magnetic multipoles and 
from the order $\eps$ for the electric ones. In the case of the harmonic time   
variation, we stand out only the parameters $d$, $c$ and $\omega$. In the case $(\mu,\eps)=
(4,5)$, considering the $c$ dependence of the coefficients given by equations \eref{anexpl},
and using equations \eref{el1}-\eref{mag1}, 
we can write
\beqa\label{p45}
&~&\fl \dot{\psftr}^{(1)}\sim a_{10}+a_{11}\frac{\omega^2d^2}{c^2}+a_{12}\frac{\omega^4
d^4}{c^4}+\dots,\;\;
 \dot{\psftr}^{(2)}\sim a_{20}d +a_{21}\frac{\omega^2d^3}{c^2}+\dots,\nonumber\\
&~&\fl \dot{\psftr}^{(3)}\sim a_{30}d^2+\frac{\omega^2d^4}{c^2}+\dots,\;\;
  \dot{\psftr}^{(4)}\sim a_{40}d^3+\dots,\;\;\dot{\psftr}^{(5)}\sim a_{50}d^4+\dots;\nonumber\\  
&~&\fl \dot{\msftr}^{(1)}\sim a'_{10}d+a'_{11}\frac{\omega^2d^3}{c^2}+\dots,\;\;
\dot{\msftr}^{(2)}\sim a'_{20}d^2+a'_{21}\frac{\omega^2d^4}{c^2}+\dots,\nonumber\\
&~&\fl\dot{\msftr}^{(3)}\sim a'_{30}d^3+\dots,\;\;\;
\dot{\msftr}^{(4)}\sim a'_{40}d^4+\dots  
  \eeqa
where $a_{ij}$ and $a'_{ij}$ are functions of time proportional with $\sin{\omega t}$ or  
$\cos{\omega t}$, and the magnitude order of a common factor is not considered. 
We point out that in equations \eref{p45} a contribution to the absolute order of magnitude 
of the final results gives the current density $j\sim v\sim \omega d$. This factor present 
in all the expressions from equations \eref{p45}, associated with $1/c$ factors from 
equations \eref{iepsmu}, \eref{Frf} and \eref{finmomf}, gives a factor $(d/\la)^2\sim
 (v/c)^2$ in the expressions of the corresponding expansions. This common factor is irrelevant 
 four our considerations.   
 If we consider 
equations \eref{p45} in the multipolar expansion \eref{iepsmu}, we can write in the 
  case $(\mu,\eps)=(4,5)$ the following result   for the relative orders of magnitude  
  of the different terms
\beqa\label{oamgn}  
&~&  {\mathcal I}^{(4,5)}\sim A_0+A_2\zeta^2+
A_4\zeta^4+A_6\zeta^6+
A_8\zeta^8,\;\;\;\zeta=d/\la.
\eeqa

From equations \eref{ForP} and \eref{ForM}, one can see that ${\mathcal I}^{(5,6)}$ 
can be written in the form of a similar expansion in which $A_0$, $A_2$, and $A_4$ 
are unchanged, and only $A_6$, $A_8$,...will be different. Consequently, for a consistent 
approximation of ${\mathcal I}$ using the multipole expansions up to $(\mu,\eps)=(4,5)$ 
one should keep only the terms of the order $(d/\la)^n$ for $n< 4$.  
\par Similar considerations may be applied for the evaluation of the recoil force and 
radiated angular momentum. From the results given in \cite{dub} and \cite{ra}, it seems 
that  the authors claim to give the first 3 terms of the $d/\la$ expansion of the 
radiated power, the first two terms of the expansion of the recoil force and again the 
first 3 terms of the expansion of the angular momentum loss.  
 Our evaluations  give for these cases, very different results. It seems that, in the 
above mentioned papers, the approximation criteria are not applied  consistently. The  
comprising of the factor $1/c$ in the definitions of different parameters as, for example, 
for the  toroid dipole, seems to be one of the sources of errors. We point out the 
benefit of using a system-free expression of Maxwell's equations, as in the present paper.
 So, the factors $c$ included in different definitions in  Gauss system of units is 
 represented here by $\al$ which is no counting to the evaluation of approximation order.
  This fact was observed also in \cite{DV} in the case of the energy loss.
\par We give bellow our results for the above mentioned approximations.
\par In the case of the radiated power, it is necessary to calculate ${\mathcal I}^{(4,5)}$ 
and retain only the powers of $\zeta$ up to four. One obtains
 
 \beqa\label{ifinal}   
&~&\fl {\mathcal I}=\frac{1}{4\pi\eps_0c^3}\left[\frac{2}{3}\ddot{\pv}^2-\frac{4}{3c^2}
\ddot{\pv}\ct\tdot{\tv}+\frac{1}{20c^2}\tdot{\pc}^{(2)}\ct\tdot{\pc}^{(2)}
+\frac{2\al^2}{3c^2}\ddot{\mv}^2+\frac{2}{3c^4}\tdot{\tv}^2+\frac{4}{3c^4}
\ddot{\pv}\ct
\ddot{\tdot{\bbox{T}}}_{(2)}\right.\nonumber\\
&~&\fl\left.-\frac{1}{10c^4}\tdot{\pc}\ct
\qdot{\Tsf}^{(2)}_{(1)}+\frac{2}{945c^4}\qdot{\pc}^{(3)}\ct\qdot{\pc}^{(3)}
+\frac{4\al^2}{3c^4}\ddot{\mv}\ct\qdot{\muv} +\frac{\al^2}{20c^4}\tdot{\mc}^{(2)}\ct
\tdot{\mc}^{(2)}\right]+O(\zeta^6).
\eeqa
 The recoil force is given by
 \beqa\label{recoil}
 &~&\fl 4\pi\eps_0\bbox{F}=-\frac{1}{c^5}\left[\frac{1}{5}\ddot{\bbox{p}}\ct
\tdot{{\mathcal P}}^{(2)}-\frac{2\al}{3}\ddot{\bbox{m}}\times\ddot{\bbox{p}}
-\frac{2\al}{3c^2}\big(\qdot{\bbox{\mu}}\times\ddot{\bbox{p}}-\ddot{\bbox{m}}
\times\tdot{\bbox{t}}\big)+\frac{\al^2}{5c^2}\ddot{\bbox{m}}\ct\tdot{
{\mathcal M}}^{(2)}\right.\nonumber\\
&~&\fl\left. -\frac{1}{5c^2}\big(\ddot{\bbox{p}}\ct\qdot{\Tsf}^{(2)}_{(1)}+
\tdot{\bbox{t}}\ct\tdot{{\mathcal P}}^{(2)}\big)+\frac{4}{315c^2}
\tdot{{\mathcal P}}^{(2)}\ct\qdot{{\mathcal P}}^{(3)}
-\frac{\al}{30c^2}\evec_i\eps_{ijk}\tdot{{\mathcal M}}_{qj}
\tdot{{\mathcal P}}_{qk}+O(\zeta^5),
\right]
\eeqa
  and the radiated angular momentum
  \beqa\label{lrad}
&~&\fl 4\pi\eps_0\frac{\rmd\bbox{L}}{\rmd t}=\frac{1}{c^3}\left[-\frac{2}{3}\dot{\pv}\times\ddot{\pv}+
\frac{2}{3c^2}\dot{\pv}\times\tdot{\tv}+\frac{2}{3c^2}\ddot{\tv}\times\ddot{\pv}
-\frac{2\al^2}{3c^2}\dot{\mv}\times\ddot{\mv}+\frac{\al}{5c^2}\ddot{\mv}\ct\ddot{\pc}^{(2)}
\right.\nonumber\\
&~&\fl \left.+\frac{\al}{5c^2}\dot{\mv}\ct\tdot{\pc}^{(2)}-\frac{\al}{5c^2}\dot{\pv}\ct
\tdot{\mc}^{(2)}-\frac{1}{10c^2}\evec_i\eps_{ijk}\ddot{\pc}_{qj}\tdot{\pc}_{qk}
-\frac{2}{3c^4}\dot{\pv}\times\ddot{\tdot{\bbox{T}}}_{(2)}-\frac{2}{3c^4}\ddot{\tv}\times
\tdot{\tv}\right.\nonumber\\
&~&\fl\left. -\frac{2\al^2}{3c^4}\tdot{\muv}\times\ddot{\mv}-\frac{2}{3c^4}
\qdot{\bbox{T}}_{(2)}\times\ddot{\pv}-\frac{2\al^2}{3c^4}\dot{\mv}\times\qdot{\muv}
+\frac{\al}{5c^4}\qdot{\muv}\ct
\ddot{\pc}^{(2)}-\frac{\al}{5c^4}\ddot{\mv}\ct\tdot{\Tsf}^{(2)}_{(1)}\right.\nonumber\\
&~&\fl \left. -\frac{\al}{5c^2}\ddot{\pv}\ct\ddot{\mc}^{(2)}+\frac{\al}{5c^4}\tdot{\tv}
\ct\ddot{\mc}^{(2)}-\frac{\al}{5c^4}\ddot{\pv}\ct\qdot{\mu}^{(2)}
-\frac{\al}{5c^4}\dot{\mv}\ct\qdot{\Tsf}^{(2)}_{(1)}+\frac{\al}{5c^4}\tdot{\muv}
\ct\tdot{\pc}^{(2)}-\frac{\al}{5c^4}\dot{\pv}\ct\ddot{\tdot{\mu}}^{(2)}\right.
\nonumber\\
&~&\fl \left. +\frac{\al}{5c^4}\ddot{\tv}\ct\tdot{\mc}^{(2)}
+\frac{1}{10c^4}\evec_i\eps_{ijk}\ddot{\pc}_{qj}\qdot{\Tsf}_{qk}
+\frac{1}{10c^4}\evec_i\eps_{ijk}\tdot{\Tsf}_{qj}\tdot{\pc}_{qk}
-\frac{\al^2}{10c^4}\evec_i\eps_{ijk}\ddot{\mc}_{qj}\tdot{\mc}_{qk}\right.\nonumber\\
&~&\fl\left. +\frac{2\al}{105c^4}\tdot{\pc}^{(3)}\ct\tdot{\mc}^{(2)}
-\frac{2\al}{105c^4}\tdot{\mc}^{(3)}\ct\tdot{\pc}^{(2)}
+\frac{2\al}{105c^4}\qdot{\pc}^{(3)}\ct\ddot{\mc}^{(2)}-\frac{2\al}{105c^4}
\qdot{\mc}^{(3)}\ct\ddot{\pc}^{(2)}\right.\nonumber\\
&~&\fl\left. -\frac{2}{315c^4}\evec_i\eps_{ijk}\tdot{\pc}_{qpj}\qdot{\pc}
_{qpk}
\right]+O(\zeta^6).
\eeqa
The following notations was introduced in the last three equations:
\beq
\fl \bbox{p},\bbox{m},\bbox{\mu},\bbox{t}\;\mbox{for}\;{\mathcal P}^{(1)},\;{\mathcal M}^{(1)},\;
\;\mu^{(1)},\Tsf^{(1)}_{(1)},\;\mbox{and}\;\bbox{T}_{(2)}=\evec_i\Tsf^{(1)}_{(2)i}
\eeq
For comparison, we give below the results from \cite{dub,ra} with our notation but 
written in Gauss unit system. So, the result from \cite{dub} is 
\footnote{For comparison with our formulae one takes $4\pi\eps_0\to 1,\; \al\to c$}

\beqa\label{dub}
&~&\fl {\mathcal I}_{DT}=\frac{2}{3c^3}\vert\ddot{\pv}-\frac{1}{c^2}\tdot{\tv}\vert^2
+\frac{2}{3c^3}\vert\ddot{\mv}\vert^2+\frac{1}{20c^5}\big(\tdot{\pc}_{ij}-\frac{1}{c^2}
\tdot{\Tsf}_{(1)ij}\big)\big(\tdot{\pc}_{ij}-\frac{1}{c^2}\tdot{\Tsf}_{(1)ij}\big)\nonumber\\
&~&\fl +\frac{2}{945c^7}\big(\qdot{\mc}_{ijk}\qdot{\mc}_{ijk}+\qdot{\mc}_{ijk}
\qdot{\mc}_{ijk}\big).
\eeqa
In this expression are present terms as  $\qdot{\mc}_{ijk}\qdot{\mc}_{ijk}$ and
$\tdot{\Tsf}_{(1)ij}\tdot{\Tsf}_{(1)ij}$ of order six in $\zeta$ and are missed 
terms of the order four as $\tdot{\pc}_{ijk}\tdot{\pc}_{ijk}$ for example. This expression 
may not to be considered as an expansion up to the order six in $\zeta$ beacause in  this 
case many terms must be added.\\
The result for ${\mathcal I}$ from \cite{ra} is given by
\beqa\label{ra}
&~&\fl {\mathcal I}_{RV}=\frac{2}{3c^3}\ddot{\pv}^2+\frac{2}{3c^5}\ddot{\mv}^2
-\frac{4}{3c^5}\ddot{\pv}\ct
\tdot{\tv}+\frac{2}{3c^7}\tdot{\tv}^2+\frac{4}{3c^7}\ddot{\mv}\ct\qdot{\muv}\nonumber\\
&~&\fl +\frac{1}{20c^7}\tdot{\pc}_{ij}\tdot{\pc}_{ij}+
\frac{1}{20c^7}\tdot{\mc}_{ij}\tdot{\mc}_{ij}.
\eeqa
In equation \eref{ra}, the terms $\tdot{\tv}^2$ and $\ddot{\mv}\ct\qdot{\muv}$ represent 
corrections of order four but many terms of the same order are missing compared with 
equation \eref{ifinal}. Similar conclusions may be formulated regarding the results for 
recoil force and angular momentum.  
 
\appendix
\section{Gauge invariant reduction of multipole Cartesian tensors} 
The transformations \eref{red} involve a sequence of operations meant to obtain 
the symmetric and traceless part of the tensors  implied. Let an $n$-rank tensor $\Lsft^{(n)}$
of magnetic type  i.e. symmetric in the first $n-1$ indices 
and verifying the property  
$\Lsf_{i_1\dots i_{k-1}\,j\,i_{k+1}\dots i_{n-1}\,j}=0,\; k=1\dots n-1$. 
The the symmetric part of this tensor is given by
\beqan
\fl &~&\Lsf_{\sym i_1\dots i_n}=\frac{1}{n}\big[\Lsf_{i_1\dots i_n}+
\Lsf_{i_ni_2\dots i_1}+\dots +\Lsf_{i_1\dots i_n\,i_{n-1}}\big]
=\Lsf_{i_1\dots i_n}-\frac{1}{n}\suml^{n-1}_{\la=1}\eps_{i_{\la}i_nq}\Ncal^{(\la)}
_{i_1\dots i_{n-1}\,q}\big[\Lsft^{(n)}\big]
\eeqan
where $\Nmc^{\dots (\la)}_{\dots}$ is the component with the index $i_{\la}$ suppressed.
\par The operator $\Ncal$  defines a correspondence between $\Lsft^{(n)}$ and 
a tensor of rank  $(n-1)$: 
\beqa\label{opN}
&~&\fl\Lsft^{(n)}\longrightarrow \Ncal\big[\Lsft^{(n)}\big]:\;
 \left[\Ncal\big[\Lsft^{(n)}\big]\right]_{i_1\dots i_{n-1}}\equiv 
 \Nmc_{i_1\dots i_{n-1}}\big[\Lsft^{(n)}\big]=
 \eps_{i_{n-1}ps}\Lsf_{i_1\dots i_{n-2}ps}
 \eeqa 
which is  fully symmetric in the first $n-2$ indices and the contractions of the last 
index with all the previous indices give null results. So, the tensor $\Ncal[\Lsft^{(n)}]$ 
is  of the type $\msft^{(n-1)}$. 
Particularly,
\beqa\label{ms2}
\fl&~&\Ncal^{\,2k}\big[\msf{(n)}\big]=\frac{(-1)^kn}{(n+1)\al}\intl_{\dom}\xi^{2k}
\xivec^{n-2k}\times\jvec\rmd^3\xi,\nonumber\\
\fl&~&\Ncal^{\,2k+1}\big[\msft^{(n)}\big]=\frac{(-1)^kn}{(n+1)\al}\intl_{\dom}
\xi^{2k}\xivec^{n-2k-1}\times(\xivec\times\jvec)\rmd^3\xi,\;k=0,1,2\dots
\eeqa
Let a fully symmetric tensor $\Ssft^{(n)}$ and the {\it detracer} operator $\Tcal$  
introduced in \cite{ap}. This operator acts on a totally symmetric tensor $\Ssft^{(n)}$ 
so that $\Tcal[\Ssft^{(n)}]$ is a fully symmetric and traceless tensor of rank $n$. 
The {\it detracer theorem}  states that \cite{ap} \footnote{In this equation, the 
definition of the symmetric and traceless part of the tensor $S^{(n)}$ differs from 
that used in \cite{ap} by a factor $1/(2n-1)!!$.}
\beqa\label{detrth}
&~&\fl\left[\Tcal\big[\Ssft^{(n)}\big]\right]_{i_1\dots i_n}=
\suml^{[n/2]}_{m=0}\frac{(-1)^m(2n-1-2m)!!}{(2n-1)!!}\delta_{\{i_1i_2}\dots
\delta_{i_{2m-1}i_{2m}}\Ssf^{(n:m)}_{i_{2m+1}\dots i_n\}}
\eeqa
where $[n/2]$ denotes the integer part of $[n/2]$,  $A_{\{i_1\dots i_n\}}$ is 
the sum over all permutations of the symbols $i_q$  which give distinct terms, and 
$\Ssf^{(n:m)}_{i_{2m+1}\dots i_n}$ denotes the components of the $(n-2m)$th-order tensor 
obtained from $\Ssft^{(n)}$ by contracting $m$ pairs of symbols $i$.
\par It is useful to introduce here another operator $\Laop$ by the relationship 
\beqa\label{defla}
&~&\fl \left[\Tcal\big[\Ssft^{(n)}\big]\right]_{i_1\dots i_n}=
\Ssf_ {i_1\dots i_n}-\delta_{\{i_1i_2}\big[\Laop\big[\Ssft^{(n)}\big]\big]
_{i_3\dots i_n\}}
\eeqa
where $\Laop\big[\Ssft^{(n)}\big]$ defines a fully symmetric tensor 
of rank $n-2$. From this definition 
together with the theorem \eref{detrth}, we obtain

\beqa\label{lamatr}
\fl\Lambda_{i_1\dots i_{n-2}}\big[\Ssft^{(n)}\big]=
\suml^{[n/2-1]}_{m=0}\frac{(-1)^m[2n-1-2(m+1)]!!}{(m+1)(2n-1)!!}
\delta_{\{i_1i_2}\dots \delta_{i_{2m-1}i_{2m}}\Ssf^{(n:\,m+1)}_{i_{2m+1}\dots i_{n-2}\}}.
\eeqa
In the following, for simplifying the notation, all arguments of the operator 
$\Laop$ should be considered as a symmetrized tensor i.e. $\Laop[\Tsft^{(n)}]=\Laop[\Tsft^{(n)}
_{\scr sym}]$ for any tensor $\Tsft^{(n)}$. The same applies to the operator 
$\Tcal$: $\Tcal[\Tsft^{(n)}]=\Tcal[\Tsft^{(n)}_{\scr sym}]$.
\par The following four transformation properties of the multipole tensors and potentials   
may be used for establishing the results from \cite{vr}.
\par I. Let the transformation of the $n$th-order magnetic tensor: 
\beqa\label{trLM}
&~&\fl \msft^{(n)}\rightarrow\msft^{(n)}_{(L)}:\;
\msf_{(L)i_1 \dots i_n}=\msf_{i_1 \dots i_n}-\frac{1}{n}\suml^{n-1}_{\la=1}
\eps_{i_{\la}i_nq}\Nmc^{(\la)}_{i_1 \dots i_{n-1}\,q}\big[\Lsft^{(n)}(t_0)].
\eeqa  
Let us substitute in the expansion of the potential $\Avec$ the tensor $\msft^{(n)}$ by 
$\msft^{(n)}_{(L)}$ obtaining 
\beqa\label{trA2}
\fl&~& \Avec[{\scriptstyle \msft^{(n)}\rightarrow\msft^{(n)}_{(L)}}]=\Avec
-\frac{\mu_0}{4\pi}\frac{(-1)^{n-1}}{n!n}\,\evec_i\,\d_j\d_{i_1}\dots\d_{i_{n-1}}
\left[\frac{1}{r}\suml^{n-1}_{\la=1}\eps_{ijk}\eps_{i_{\la}kq}
\Nmc^{(\la)}_{i_1\dots i_{n-1}q}\big[\Lsft^{(n)}(t_0)\big]\right]\nonumber\\
\fl&~&=\Avec
+\frac{\mu_0}{4\pi}\frac{(-1)^{n-1}}{n!n}\,\evec_i\,
\d_j\d_{i_1}\dots\d_{i_{n-1}}
\frac{1}{r}\big[\big(\delta_{i\,i_1}\Nmc_{i_2\dots i_{n-1}\,j}+\dots\delta_{i\,i_{n-1}}
\Nmc_{i_1\dots i_{n-2}\,j}\big)\nonumber\\
\fl&~&- \big(\delta_{j\,i_1}\Nmc_{i_2\dots i_{n-1}\,i}+\dots\delta_{j\,i_{n-1}}
\Nmc_{i_1\dots i_{n-2}\,i}\big)\big]\big[\Lsft^{(n)}(t_0)\big]\nonumber\\
\fl&~&=\Avec+\nablav\Psi(\rvec,t)
+\frac{\mu_0}{4\pi}\frac{(-1)^{n}(n-1)}{n!c^2n}\nablav^{n-2}\ct \left[\frac{1}{r}
\ddot{\Ncal}\big[\Lsft^{(n)}(t_0)\big]\right].
\eeqa   
Here the equation   
  $ [\Delta-(1/c^2)\d^2/\d t^2][f(t_0)/r]=0,\;\;r\,\ne\, 0$ is  considered.
  The function $\Psi$ is a solution of the homogeneous wave equation for $r\neq 0$ 
  and the corresponding  expression is irrelevant. 
 Let the transformation
 \beqa\label{1}
 &~&\fl
 \psft^{(n-1)}\rightarrow\psft^{'(n-1)}=\psft^{(n-1)}+a_1(n)\dot{\Ncal}
 \big[\Lsft^{(n)}\big],\;\;\;\;a_1(n)=-\frac{\al}{c^2}\frac{n-1}{n^2}.
 \eeqa
 Introducing the transformed potentials produced by the substitution 
 $ \psft^{(n-1)}\rightarrow\psft^{'(n-1)}$, we obtain 
\beqan
\fl&~& \Avec[{\scriptstyle \msft^{(n)}\rightarrow\msft^{(n)}_{(L)},\;
\psft^{(n-1)}\rightarrow\psft^{'(n-1)}}]=\Avec+\nablav\Psi,
\;\;\;\;\Phi[{\scriptstyle \psft^{(n-1)}\rightarrow\psft^{\,'(n-1)}}]=\Phi
-\frac{\d \Psi}{\d t}.
\eeqan
So, the transformation \eref{trLM} produces changes in the potentials which, up to 
a gauge transformation, are compensated by the transformation \eref{1}.  	
\par II. Let the transformation of the $n$th-order electric tensor:
\beqa\label{trLP}
&~&\fl \psft^{(n)}\rightarrow\psft^{(n)}_{(L)}:\;
\psf_{(L)i_1 \dots i_n}=\psf_{i_1 \dots i_n}-\frac{1}{n}\suml^{n-1}_{\la=1}
\eps_{i_{\la}i_nq}\Nmc^{(\la)}_{i_1 \dots i_{n-1}\,q}\big[\Lsft^{(n)}(t_0)].
\eeqa  
We obtain
\beqan
\fl&~& \Avec[{\scriptstyle \psft^{(n)}\rightarrow\psft^{(n)}_{(L)}}]=\Avec
+\frac{\mu_0}{4\pi}\frac{(-1)^{n-1}(n-1)}{n!\,n}\nablav\times\left\{\nablav^{n-2}\ct
\left[\frac{1}{r}\dot{\Ncal}\big[\Lsft^{(n)}\big]\right]\right\},
\;\;\;\Phi[{\scriptstyle \psft^{(n)}\rightarrow\psft^{(n)}_{(L)}}]=\Phi.
\eeqan
The change of the vector potential  is compensated by the transformation
\beqa\label{2}
&~&\fl 
\msft^{(n-1)}\longrightarrow \msft^{(n-1)}+a_2(n)\dot{\Ncal}\big[\Lsft^{(n)}\big],\;\;
\;\;\;a_2(n)=\frac{n-1}{\al n^2}=-\frac{c^2}{\al^2}a_1(n).
\eeqa
\par III. Let the transformation of the magnetic vector of rank $n$:
\beqa\label{trsM}
&~&\fl\msft^{(n)}\longrightarrow \msft^{(n)}_{\scriptstyle (S)}:\;\;
\msft_{{\scriptstyle (S)}i_1\dots i_n}=\msft_{i_1\dots i_n}-\suml_{D(i)}\delta_{i_1i_2}
\Lambda_{i_3\dots i_n}\big[\Ssft^{(n)}(t_0)\big]
\eeqa
where $\Ssft^{(n)}$ is a fully symmetric tensor.
The alteration of the vectorial potential is eliminated by the transformation
\beqa\label{3}
&~&\fl\msft^{(n-2)}\longrightarrow\msft^{(n-2)}+b(n)\ddot{\bbox{\Lambda}}\big[\Ssft^{(n)}\big],
\;\;\;\;\;b(n)=\frac{n-2}{2c^2n}.
\eeqa 
\par IV. The transformation
\beqa\label{trsP}
&~&\fl\psft^{(n)}\longrightarrow \psft_{\scriptstyle (S)}:\;\;
\psft_{{\scriptstyle (S)}i_1\dots i_n}=\psft_{i_1\dots i_n}-\suml_{D(i)}\delta_{i_1i_2}
\Lambda_{i_3\dots i_n}\big[\Ssft^{(n)}\big]
\eeqa
produces the following changes of the potentials:
\beqan
\fl&~&\Avec[{\scriptstyle \psft^{(n)}\rightarrow \psft^{(n)}_{(S)}}]=
\Avec-\frac{\mu_0}{4\pi}\frac{(-1)^{n-1}}{n!}\evec_i\d_{i_1}\dots\d_{i_{n-1}}
\left[\frac{1}{r}\suml_{D(i)}\delta_{i_1i_2}\dot{\Lambda}_{i_3\dots i_n}
\big[\Ssft^{(n)}\big]\right]\\
\fl&~&=\Avec+\nablav \Psi'-\frac{\mu_0}{4\pi}\frac{(-1)^{n-1}(n-2)(n-1)}{2n!c^2}
\nablav^{n-3}\ct\left[\frac{1}{r}\tdot{\bbox{\Lambda}}\big[ \Ssft^{(n)}\big]
\right]
\eeqan	
with $\Psi'$, as $\Psi$, satisfying the homogeneous wave equation 
and
\beqan
\fl&~&\Phi\big[{\scriptstyle \psft^{(n)}\rightarrow\psft^{(n)}_{(S)}}\big]=
\Phi+\frac{\mu_0}{4\pi}\frac{(-1)^{n-1}n(n-1)}{2c^2n!}\nablav^{n-2}\ct\left[\frac{1}{r}
\ddot{\bbox{\Lambda}}\big[\Ssft^{(n)}\big]\right].
\eeqan
Let the transformation 
\beqa\label{4}
&~&\fl 
\psft^{(n-2)}\longrightarrow \psft"^{(n-2)}=\psft^{(n-2)}+b(n)\ddot{\bbox{\Lambda}}
\big[\Ssft(t_0)\big]
\eeqa
with $b(n)$ given by equation \eref{3}. The effect of the transformation \eref{4} 
on the potential $ \Avec$ is the compensation of the extra-gauge term . So
$\Avec\big[{\scriptstyle \psft^{(n)}\rightarrow\psft^{(n)}_{(S)},
\psft^{(n-2)}\rightarrow\psft^{"(n-2)}}\big]=\Avec+\nablav\Psi'$
but it is easy to see that the modification of the scalar potential $\Phi$ produced 
by the transformation \eref{4} together with the modification due to the transformation 
\eref{trsP} give
$\Phi\big[{\scriptstyle \psft^{(n)}\rightarrow\psft^{(n)}[\Ssft^{(n)}],
\psft^{(n-2)}\rightarrow\psft^{"(n-2)}}\big]=\Phi-\d\Psi'/\d t$
the total effect of the transformations \eref{trsP} and \eref{4} being a gauge transformation 
of the potentials.  


\par 4.  Let the gauge invariant process of reducing the multipole tensors begin from the order 
$n=\eps$ in the case of 
the electric tensors  and from $n=\mu$ for the magnetic ones, and go downward up to 
lowest value, $n=1$. Generally, 
we may choose $\eps > \mu$ as will be seen in the following.  
In \cite{jpa04} are given the formulae for the results $\widetilde{\psf}^{n}$ and 
 $\widetilde{\msf}^{n}$:

\beqa\label{ForP}
\fl\widetilde{\psft}^{(n)}={\mathcal P}^{(n)}&+&{\mathcal T}\left\{
\suml^{[(\eps-n)/2]}_{k=1}A^{(n)}_k\frac{\rmd^{2k}}{\rmd t^{2k}}\Lambda^k\big[
\psft^{(n+2k)}\big]\right.\nonumber\\
\fl&~&+\left.\suml^{[(\mu-n-1)/2]}_{k=0}\frac{\rmd^{2k+1}}{\rmd t^{2k+1}}\suml^k_{l=0}
B^{(n)}_{kl}\latens^l\Ncal^{2k-2l+1}\big[\msft^{(n+1+2k)}\big]\right\},
\eeqa
\beqa\label{ForM}
\fl&~&\widetilde{\msft}^{(n)}={\mathcal M}^{(n)}+{\mathcal T}\left\{
\suml^{[(\mu-n)/2]}_{k=1}\frac{\rmd^{2k}}{\rmd t^{2k}}\suml^{k}_{l=0}
C^{(n)}_{kl}\latens^l\Ncal^{2k-2l}\big[\msft^{(n+2k)}\big]\right\}
\eeqa
where
\beqa\label{AB}
\fl&~& A^{(n)}_k=\prod^{k}_{l=1}b(n+2l),\nonumber\\
\fl&~&B^{(n)}_{kl}=\prod^l_{q=1}b(n+2q)\prod^{k-l}_{h=0}a_1(n+1+2k-2h)\prod^{k-l-1}
_{s=0}a_2(n+2k-2s),
\eeqa

\beqa\label{C}
\fl&~&C^{(n)}_{kl}=\prod^l_{q=1}b(n+2q)\prod^{k-l-1}_{h=0}a_1(n+2k-2h)\prod^{k-l-1}_{s=0}
a_2(n-1+2k-2s)
\eeqa
and
\beqa\label{an}
&~&\fl a_1(n)=-\frac{\al}{c^2}\,\frac{n-1}{n^2},\;\;a_2(n)=\frac{1}{\al}\frac{n-1}{n^2}=
-\frac{c^2}{\al^2}\,a_1(n),\;\;b(n)=\frac{n-2}{2c^2n}.
\eeqa
By introducing equations \eref{an} in equations \eref{AB} and \eref{C}, we obtain
\beqa\label{anexpl}
&~&\fl A^{(n)}_k=\frac{1}{2^kc^{2k}}\frac{n}{n+2k},\nonumber\\
&~&\fl B^{(n)}_{k,l}=\frac{(-1)^{k-l+1}\al}{2^lc^{2k+2}}\frac{n(n+2l)!}{(n+2k+1)(n+2k+1)!},\nonumber\\
&~&\fl C^{(n)}_{k,l}=\frac{(-1)^{k-l}}{2^lc^{2k}}\frac{n(n+2l)!}{(n+2k)(n+2k)!}.
\eeqa
 ${\mathcal P}^{(n)}$ and ${\mathcal M}^{(n)}$ stand for the "static" expressions of 
the reduced multipole tensors:
\beqan
\fl&~& {\mathcal P}^{(n)}(t)={\mathcal T}[\psft^{(n)}]=\frac{(-1)^n}{(2n-1)!!}\intl_{\mathcal D}
\rho(\rvec,t)r^{2n+1}\nablav^n\frac{1}{r}\rmd^3x,\\
\fl&~&{\mathcal M}^{(n)}(t)={\mathcal T}[\msft^{(n)}]=\frac{(-1)^n}{\al (n+1)(2n-1)!!}\suml^n_{\la=1}
\intl_{\mathcal D}r^{2n+1}\big[\jvec(\rvec,t)\times\nablav\big]_{i_{\la}}\d^{(\la)}
_{i_1\dots i_n}\frac{1}{r}\rmd^3x.
\eeqan
In  formulae above one should consider that 
$\prod^L_{k=l} F_k=1 \;\mbox{if}\;L\,<\,l.$
  In the case $\eps=\mu+1$, we may write 
 \beqa\label{grouping}
\fl&~&\widetilde{\psft}^{(n)}={\mathcal P}^{(n)}+
\suml^{[(\eps-n)/2]}_{k=1}\frac{(-1)^k}{c^{2k}}\frac{\rmd^{2k-1}}{\rmd t^{2k-1}}
\Tsf^{(n)}_{(k)},\nonumber\\
\fl&~&\Tsf^{(n)}_{(k)}=(-1)^kc^{2k}{\mathcal T}\left[A^{(n)}_k\Lambda^k\big[\dot{\psf}^{(n+2k)}\big]+
\suml^{k-1}_{l=0}B^{(n)}_{k-1,\,l}\Lambda^l{\mathcal N}^{2k-2l-1}\big[\msf^{(n+2k-1)}\big]
\right]\ ,
\eeqa
where  besides the usual electric and magnetic multipole moments, a third 
multipole family, the toroid moments and, generally,  mean-square radii of various orders,
 are involved \cite{dub,ra}.
\par For applications, we we give the results of the gauge invariant reduction 
of the electric and magnetic multipole tensors, begining the reduction procedure from 
the rank $\eps=5$ for the electric tensors, and from $\mu=4$ for the magnetic ones.
From equation \eref{grouping} we have for the electric multipole tensors,
\beqa\label{el1}
&~&\fl \widetilde{\psft}^{(1)}={\mathcal P}^{(1)}-\frac{1}{c^2}\dot{\Tsft}^{(1)}_{(1)}+
\frac{1}{c^4}\tdot{\Tsf}^{(1)}_{(2)},\nonumber\\
&~&\fl \big[\Tsf^{(1)}_{(1)}\big]_i=-c^2{\mathcal T}\left\{A^{(1)}_1\Lambda\big[\dot{\psft}^{(3)}\big]+
B^{(1)}_{0,0}{\mathcal N}\big[\msft^{(2)}\big]\right\}_i
=\frac{1}{10}\intl_{\dom}\big[(\xivec\cdot\jvec)\xi_i-2\xi^2j_i\big]\rmd^3\xi,\nonumber\\
&~&\fl \big[\Tsf^{(1)}_{(2)}\big]_i=
c^4{\mathcal T}\left\{A^{(1)}_2\Lambda^2\big[\dot{\psft}^{(5)}\big]+B^{(1)}_{1,0}{\mathcal 
N}^3\big[\msft^{(4)}\big]+B^{(1)}_{1,1}\Lambda{\mathcal N}\big[\msft^{(4)}\big]\right\}_i\nonumber\\
&~&\fl =-\frac{1}{280}\intl_{\dom}\big[2\xi^2(\xivec\cdot\jvec)\xi_i
-3\xi^4j_i\big]\rmd^3\xi, 
\eeqa
\beqa\label{el2}
&~&\fl \widetilde{\psft}^{(2)}={\mathcal P}^{(2)}-\frac{1}{c^2}\dot{\Tsf}^{(2)}_{(1)},\nonumber\\
&~&\fl \big[\Tsf^{(2)}_{(1)}\big]_{ik}=-c^2{\mathcal T}\left\{A^{(2)}_1\Lambda\big[\dot{\psft}^{(4)}
\big]+B^{(2)}_{0,0}{\mathcal N}\big[\msft^{(3)}\big]\right\}_{ik}\nonumber\\
&~&\fl=\frac{1}{42}\intl_{\dom}\left[ 4(\xivec\cdot\jvec)\xi_i\xi_k-5\xi^2(
\xi_{ i}j_{k}+\xi_kj_i)+2\xi^2(\xivec\cdot\jvec)\delta_{ik}\right]\rmd^3\xi,
\eeqa
\beqa\label{el3}
&~&\fl  \widetilde{\psft}^{(3)}={\mathcal P}^{(3)}-\frac{1}{c^2}\dot{\Tsf}^{(3)}_{(1)},\nonumber\\
&~&\fl \big[\Tsf^{(3)}_{(1)}\big]_{ikl}=-c^2{\mathcal T}\left\{A^{(3)}_1\Lambda\big[\dot{\psft}^{(5)}
\big]+B^{(3)}_{0,0}{\mathcal N}\big[\msft^{(4)}\big]\right\}_{ikl}\nonumber\\
&~&\fl =\frac{1}{60}\intl_{\dom}\left[5(\xivec\cdot\jvec)\xi_i\xi_k\xi_l
-5\xi^2\xi_{\{i}\xi_kj_{l\}}+\xi^2(\xivec\cdot\jvec)\delta_{\{ik}\xi_{l\}}+\xi^4
\delta_{\{ik}j_{l\}}\right]\rmd^3\xi,
\eeqa
\beqa\label{el45}
&~&\fl \widetilde{\psft}^{(4)}={\mathcal P}^{(4)},\;\;\;
\widetilde{\psft}^{(5)}={\mathcal P}^{(5)}.
\eeqa
The magnetic multipole tensors are given by
\beqa\label{mag1}
&~&\fl \widetilde{\msft}^{(1)}_i={\mathcal M}^{(1)}_i+{\mathcal T}
\left\{C^{(1)}_{1,0}{\mathcal N}^2\big[\ddot{\msft}^{(3)}\big]+
C^{(1)}_{1,1}\Lambda\big[\ddot{\msft}^{(3)}\big]\right\}_i
 =m_i+\frac{1}{c^2}\ddot{\mu}_i,\nonumber\\
&~&\fl \widetilde{\msft}^{(2)}_{ik}={\mathcal M}^{(2)}_{ik}+{\mathcal T}\left\{
C^{(2)}_{1,0}{\mathcal N}^2\big[\ddot{\msft}^{(4)}\big]+C^{(2)}_{1,1}\Lambda\big[
\ddot{\msft}^{(4)}\big]\right\}_{ik}
={\mathcal M}^{(2)}_{ik}+\frac{1}{c^2}\ddot{\mu}_{ik}\nonumber\\
&~&\fl \widetilde{\msft}^{(3)}={\mathcal M}^{(3)},\;\;
\widetilde{\msft}^{(4)}={\mathcal M}^{(4)}.
\eeqa
where
\beqa\label{muimuik}
&~&\fl m_i={\mathcal M}^{(1)}_i,\;\;\mu_i=\frac{1}{20 \al }\intl_{\dom}\xi^2
(\xivec\times\jvec)_i\rmd^3\xi,\;\;\mu_{ik}=\frac{1}{42 \al}
\intl_{\dom}\xi^2\,\xi_{\{i}(\xivec\times \jvec)_{k\}}
\rmd^3\xi
\eeqa
\vspace{2.5cm}

\par 
\end{document}